\documentclass[11pt,a4paper]{article}
\usepackage[margin=25mm]{geometry}
\usepackage{amsmath,amssymb}
\usepackage[numbers]{natbib}
\let\reviewBibitemNoStop\bibitemNoStop
\renewcommand{\bibitemNoStop}{\unskip\reviewBibitemNoStop}
\usepackage{hyperref}
\usepackage{enumitem}
\usepackage{graphicx}
\usepackage{placeins}
\usepackage{tikz}
\usetikzlibrary{arrows.meta,positioning,calc,decorations.pathmorphing,decorations.markings}
\usepackage{array,booktabs}
\setlist[itemize]{leftmargin=*,itemsep=3pt,topsep=4pt}

\title{Colours and Flavours\\\large Ropes and colour reconnection as string multiplets}
\author{Christian Bierlich\\\small Department of Physics, Lund University}

\begin{document}
\maketitle

\begin{abstract}
The increase of strange-hadron yields with event activity and the enhanced charm-baryon fractions in hadronic collisions motivate a microscopic examination of how the colour environment affects hadronization. We review and develop the string description of these effects, starting from the Lund fragmentation baseline in electron--positron annihilation and following colour correlations beyond the leading-colour approximation. Few-parton examples show how junction formation and rope hadronization arise from the same SU(3) colour structure, connecting baryon formation and the screening forces that govern strange-quark production. For nuclear collisions, we introduce a local escape criterion and a phenomenological saturation of the breakup force, motivated by a three-string example of screening through background colour correlations. With common settings, the combined prescription gives a satisfactory description of light-flavour ratios in pp and p--Pb and provides O--O predictions. Charm ratios reveal a substantial remaining deficit in $\Xi_c$ production, motivating separate tests of the fragmentation baseline and its extrapolation with activity.
\end{abstract}

\section{Introduction}
High-multiplicity proton--proton collisions produce more strange hadrons per pion than low-multiplicity events, with the largest relative increase for hadrons containing several strange quarks~\cite{ALICE:2016fzo,ALICE:2019avo}. Related trends occur in proton--nucleus collisions~\cite{ALICE:2013wgn}. Strangeness enhancement was proposed as a signature of quark--gluon plasma formation~\cite{Rafelski:1982pu}, and its observation in small systems raises the question of how hadronization changes with the collision environment. The larger charm-baryon fractions measured in hadronic collisions relative to electron--positron annihilation show that the environment also affects which hadron forms around an existing heavy quark~\cite{ALICE:2020wfu,ALICE:2023sgl,ALICE:2024ocs}.

Azimuthal anisotropies, established as probes of collective expansion in heavy-ion collisions~\cite{Ollitrault:1992bk,STAR:2000ekf}, have also been observed in pp~\cite{CMS:2010ifv,Li:2012hc} and p--Pb~\cite{CMS:2012qk,ALICE:2012eyl,ATLAS:2012cix}. Multiparticle measurements showed that these correlations extend beyond particle pairs~\cite{CMS:2015yux,CMS:2016fnw,ATLAS:2017rtr}. Reviews discuss the experimental findings~\cite{Christiansen:2024bhe,ALICE:2022wpn,Adolfsson:2020dhm,Grosse-Oetringhaus:2024bwr} and their interpretation through collective expansion, initial-state correlations and kinetic or core--corona evolution~\cite{Nagle:2018nvi,Schlichting:2016sqo,Kurkela:2018ygx,Kanakubo:2019ogh}. In this paper, we review and develop the formalism describing how microscopic colour structure alters QCD string fragmentation, starting from the electron--positron baseline. We then extend the calculation through pp to p--Pb and O--O, compare with data across systems and provide predictions for unmeasured systems.

Event generators connect hard scattering and parton showers to explicit models of hadron formation~\cite{Buckley:2011ms,Campbell:2022qmc,Bierlich:2022pfr,Bellm:2015jjp,Sherpa:2019gpd}. In the Lund picture, overlapping strings can exchange colour, combine their fields and change the forces governing string breaking. String interactions can also generate transverse pressure and long-range correlations~\cite{Bierlich:2017vhg}. The sign of the elliptic response can help identify the interacting degrees of freedom~\cite{Bierlich:2024lmb}. 

\begin{figure}[t]
\centering
\includegraphics[width=\linewidth]{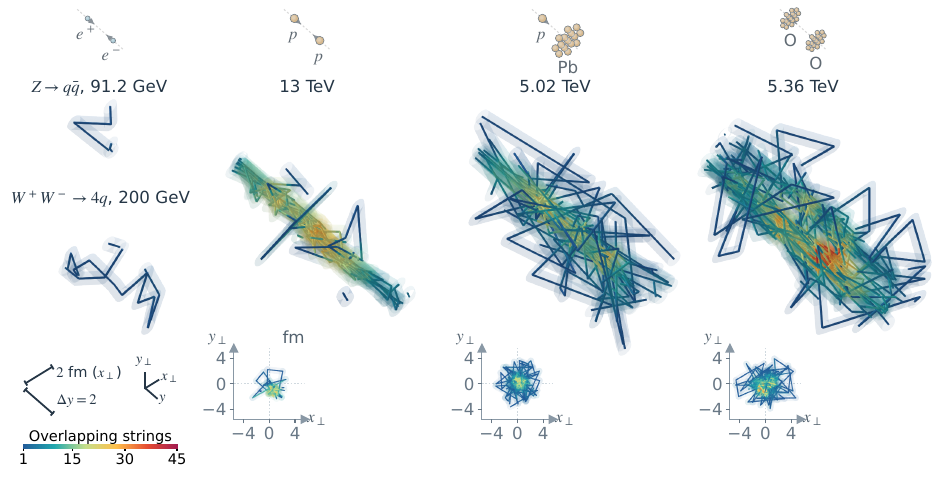}
\caption{String configurations based on selected PYTHIA and Angantyr events in five collision systems. The long axis is rapidity, and transverse positions are in fm. The colour scale gives the number of string segments within a transverse distance of $1$ fm at the same rapidity, including the displayed segment. Translucent tubes have radius $0.5$ fm. The lower panels project the displayed pp, p--Pb and O--O strings over $|y|<6$, retaining their original overlap colours.}
\label{fig:systems}
\end{figure}

Parton showers usually assign definite colour connections in the leading-colour approximation, discarding interference terms suppressed by inverse powers of the number of colours, $N_c$. At $N_c=3$, nearby charges can combine into several SU(3) representations. Colour-reconnection models use these combinations to form junctions that allow baryon production~\cite{Christiansen:2015yqa}. Higher representations in turn produce a stronger field, colour ropes, whose breakup favours strange-quark production~\cite{Bierlich:2014xba}. Colour reconnection with junction formation and rope hadronization are usually presented as separate models. Both build on the same SU(3) colour structure. The resulting flavour ratios depend on which fields overlap and how they fragment.

The electron--positron events on the left of Fig.~\ref{fig:systems} begin with a few colour charges and have no connections to beam remnants. A hadronic $Z$ decay starts with one quark--antiquark singlet, giving the baseline for ordinary string fragmentation. Hadronic $W^+W^-$ decays supply two independent singlets already at Born level. Comparing flavour yields in events with one or two hadronically decaying $W$ bosons tests interactions between the strings~\cite{Sjostrand:1993hi,Christiansen:2015yca}. Electron--positron measurements constrain the fragmentation parameters before extrapolation to hadronic collisions.

The pp, p--Pb and O--O events in Fig.~\ref{fig:systems} show dense regions of overlapping strings. The transverse projections show how the spatial extent changes across the systems. In the string description, the progression from pp to AA is mostly undramatic. Two features, however, distinguish the systems. First, nuclear systems are larger. A longer path through overlapping fields can keep a string piece with modest transverse momentum in the surrounding field until it hadronizes. Second, nuclear collisions can reach regions containing many overlapping strings. Colour correlations within that background can screen the field acting on a hadronizing string and moderate its breakup force. This motivates a phenomenological extension to p--Pb and O--O with a local escape criterion and a saturation of the breakup force in a dense coloured background. 

Section~\ref{sec:baseline} introduces the Lund string and its flavour-production rule, with a comparison to $Z$-pole data. Section~\ref{sec:colourtheory} develops the few-parton examples, their junction and rope channels, and the experimental opportunity in $WW$ events. Section~\ref{sec:collisions} describes the implementation and nuclear extension, followed by the flavour comparisons and predictions.

\section{The basic Lund string}\label{sec:baseline}
The basic structure of a simulated event, leading up to the hadronization of a Lund string, is the following: A hard process produces partons, and a parton shower resolves further radiation down to a cutoff. The event generator passes their momenta, flavours and colour connections to hadronization. In a leading-colour description those connections select ordered chains; string fragmentation converts them into hadrons~\cite{Buckley:2011ms,Bierlich:2022pfr,tHooft:1973alw}. Historically, electron--positron annihilation has supplied a clean starting point because the incoming particles carry no colour and there are no beam-remnant connections.

The starting point of the Lund string is a relativistic flux tube between separating colour charges. Its approximately linear confining potential motivates an energy per unit length $\kappa$, of order $1\,\mathrm{GeV/fm}$, while the fragmentation model supplies the dynamical prescription for producing hadrons~\cite{Andersson:1983ia,Bierlich:2022pfr}. A full hadronization model requires a dynamical fragmentation prescription in addition to the confining potential. Lattice studies resolve a transverse flux profile with a finite width~\cite{Cea:2014uja}.

The parton shower delivers an approximation to the full cross section of a multi-parton final state~\cite{Buckley:2011ms,Bierlich:2022pfr}. One of the approximations made is the so-called ``leading colour approximation'', i.e. performing the colour expansion to leading order. Here, a gluon is represented by a colour--anticolour pair and becomes a kink on an ordered string. The resulting planar connections are a practical use of large-$N_c$ organization~\cite{tHooft:1973alw}. A newly created $q\bar q$ pair terminates two neighbouring pieces; repeated breaking builds colour-singlet hadrons while sharing the available energy and momentum. Requiring consistency between fragmentation from opposite ends leads to the familiar Lund symmetric light-cone distribution, $f(z)\propto z^{-1}(1-z)^a\exp(-bm_{\perp,H}^2/z)$~\cite{Andersson:1983ia}, where $z$ is the fraction of remaining light-cone momentum taken away by the hadron, $m_{\perp,H}=\sqrt{m_H^2+\boldsymbol p_{\perp,H}^2}$ is the transverse mass of the hadron and $a$ and $b$ are parameters.

For an isolated string, transverse structure can largely be absorbed into effective parameters. Data constrain flavour suppression, spin weights and baryon production beyond the information supplied by colour counting. LEP and SLD fragmentation measurements therefore remain central constraints on hadron-collision models; the Monash tune illustrates how such constraints are combined with hadron-collider data~\cite{Skands:2014pea}.

\subsection{From a constant longitudinal force to flavour production}
\label{sec:wkb}
The dependence of string breaking on the local force follows from a semiclassical tunnelling argument~\cite{Andersson:1983ia,Schwinger:1951nm}. Work in the rest frame of a straight field segment with constant longitudinal force $F>0$, neglecting transverse gradients over the tunnelling region. For a fundamental string $F=\kappa$; the corresponding rope force will be identified in Section~\ref{sec:ropescreening}.

Transverse momentum $\mathbf k_\perp$ is conserved during tunnelling, so the longitudinal motion has effective mass $m_\perp=\sqrt{m^2+\mathbf k_\perp^2}$, where $m$ is now an effective quark mass used in fragmentation. With potential energy $V(z)=-Fz$, the relativistic energy--momentum relation gives
\begin{equation}\label{eq:dispersion}
 (E-V(z))^2=p_z^2+m_\perp^2,
 \qquad p_z^2(z)=(E+Fz)^2-m_\perp^2.
\end{equation}
The turning points satisfy $E+Fz_\pm=\pm m_\perp$. Between them $p_z^2<0$: a larger transverse mass widens the forbidden region, while a stronger force narrows it.

A WKB amplitude goes as $\exp[-\int|p_z|\,\mathrm dz]$. Squaring it gives the probability exponent
\begin{align}
 \mathcal S(F,m_\perp)
 &=2\int_{z_-}^{z_+}\sqrt{m_\perp^2-(E+Fz)^2}\,\mathrm dz\\
 &=\frac{2}{F}\int_{-m_\perp}^{m_\perp}\sqrt{m_\perp^2-u^2}\,\mathrm du,
 \qquad u=E+Fz.
\end{align}
The integral is the area of a semicircle, giving $\mathcal S=\pi m_\perp^2/F$. Thus the local pair-production weight is
\begin{equation}\label{eq:tunnelling}
 \frac{\mathrm dP_f}{\mathrm d^2k_\perp}
 \propto \exp\left(-\frac{\pi m_f^2}{F}\right)
          \exp\left(-\frac{\pi\mathbf k_\perp^2}{F}\right).
\end{equation}
For fixed effective masses and equal flavour prefactors, transverse integration gives the overall strangeness suppression factor
\begin{equation}
 \rho_s(F)=\frac{P_s(F)}{P_u(F)}
 =\exp\left[-\frac{\pi(m_s^2-m_u^2)}{F}\right].
\end{equation}
Writing $F=h\kappa$ and fixing the baseline $\rho_0=\rho_s(\kappa)$ then gives
\begin{equation}
\label{eq:rescaling}
 \rho_s(h)=\rho_0^{1/h}.
\end{equation}

When the force varies, the flavour yields must be summed over breakup environments before taking their ratio. We note that the charm quark mass is so heavy that the analogous suppression factor for charm would be $\mathcal{O}(10^{-14})$, using a constituent charm mass at ordinary string tension, so charm-pair creation is neglected in fragmentation.

\subsection{The \texorpdfstring{$Z$}{Z}-pole fragmentation baseline}
\begin{figure}[htbp]
\centering
\includegraphics[width=\linewidth]{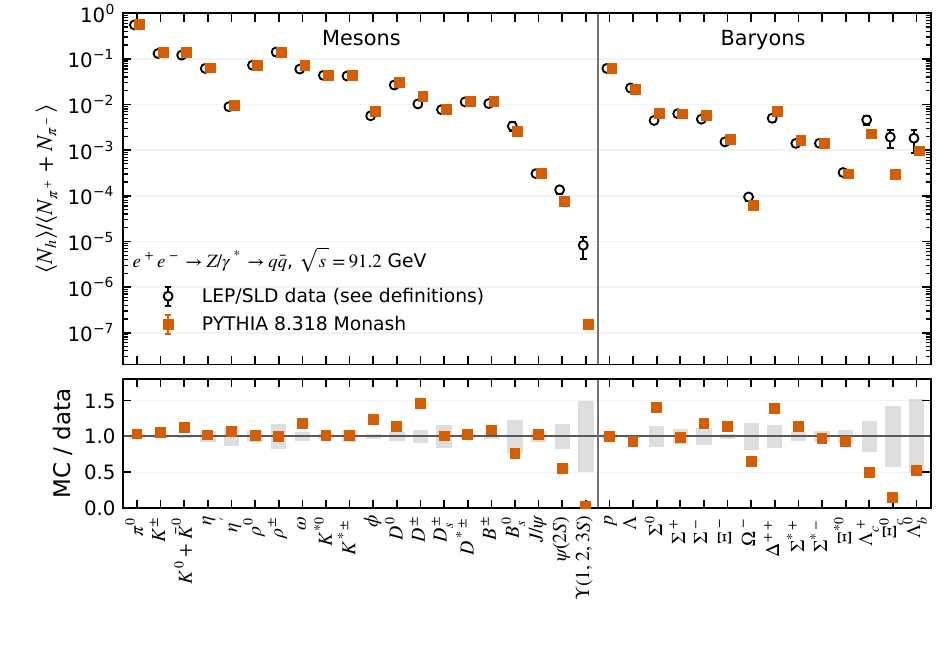}
\caption{Inclusive hadron yields relative to charged pions in hadronic $Z$ decays at $\sqrt{s}=91.2$ GeV: LEP/SLD measurements (open circles) and PYTHIA 8.319 with the unmodified Monash tune~\cite{Skands:2014pea} (squares, $10^7$ events). Charge conjugates are summed; heavy-hadron yields include bottom feed-down.}
\label{fig:zbaseline}
\end{figure}

Hadronic $Z$ decays constrain the flavour, spin and baryon weights of ordinary fragmentation~\cite{Andersson:1983ia,Skands:2014pea}. Determining model parameters from data is known as ``tuning''. The guiding principle is to use the simplest collision system that constrains a given parameter: what can be fixed in electron--positron collisions should be fixed there before turning to pp data. The comparison in Fig.~\ref{fig:zbaseline} tests a fixed tune before introducing interactions between overlapping strings.\footnote{This figure, and all other figures in this manuscript, are produced with PYTHIA 8.318 including modifications which will be published from version 8.319.} Most light-meson and ground-state baryon ratios (the PYTHIA implementation of the Lund string only produces ground-state ($L=0$) baryons, and only ground-state ($L=0$) mesons are enabled by default) are reasonably reproduced, establishing a useful reference for ordinary fragmentation, but the agreement is species dependent. 
A sub-optimal tune or a deficiency in the baseline model will carry over to predictions for larger colliding systems. In those cases it will not be clear whether a discrepancy in the larger system is due to a fault in the multi-string model, or if the baseline instead should be modified. A test of a multistring model should therefore establish the baseline for each measured species. The following discrepancies stand out:
\begin{itemize}
    \item The $\phi$ ratio is high. The $\phi$ has a special role for strangeness production, since, as the vector mixing angle is very close to ideal mixing, a $\phi$ mostly requires two subsequent strange quark breaks. This gives rise to specialized $\phi$-triggered observables \cite{Bierlich:2024jvb} but care must be taken to include the overestimated baseline.
    \item The $D^\pm$ ratio is high with both the $\Lambda_c$ and $\Xi_c$ rates low. This indicates that charm baryon rates are generally underestimated. Since charm-pair creation is neglected in fragmentation, an increase in baryonic charm would reduce the number of $D$ mesons, even in the baseline.
    \item Quarkonium ratios, $\psi(2S)$ and $\Upsilon(1S,2S,3S)$, are low. No additional quarkonium-production processes are enabled in this sample, so these discrepancies also probe the production mechanisms included in the setup. Studies of direct production in the shower have been performed \cite{Cooke:2023ldt}, but no full tune including such models exists.
    \item The $\Delta^{++}$ ratio is high, while the $\Omega^-$ rate is low. Studies exist linking this to hyperfine splitting effects arising from the mass difference between the $u,d$-type quarks and $s$-quarks \cite{Bierlich:2022vdf}, but again no full tune exists.
\end{itemize}

\section{Microscopic colour structure beyond independent strings}\label{sec:colourtheory}

\subsection{From an isolated string to interacting colour systems}
\label{sec:crworked}
Return to the hadronic $Z$ decays used to establish the fragmentation baseline. At Born level, $Z\to u_1\bar u_2$ supplies a quark--antiquark colour singlet and a single string.

A quark has three independent colour states, conventionally called red, green and blue. An SU(3) colour transformation mixes these states. The three-dimensional space on which it acts is the fundamental representation, denoted $3$ and called a triplet. An antiquark belongs to the conjugate representation $\bar3$, the antitriplet: its three basis states are antired, antigreen and antiblue.

For a quark and antiquark there are nine product states. These can be reorganized into sets that transform separately under colour rotations, the irreducible representations or multiplets:
\begin{equation}\label{eq:qqbarrepresentations}
 3\otimes\bar3=1\oplus8.
\end{equation}
Here $\otimes$ combines the two colour spaces, while $\oplus$ separates the resulting space into a one-dimensional singlet and an eight-dimensional octet. The singlet is the normalized combination
\begin{equation}\label{eq:coloursinglet}
 |1\rangle=\frac{|r\bar r\rangle+|g\bar g\rangle+|b\bar b\rangle}{\sqrt3}.
\end{equation}
It is unchanged by a simultaneous colour rotation of the quark and antiquark. The eight orthogonal combinations form the octet, which transforms like a gluon. A fixed red--antired product has both singlet and octet components. Calling the $Z$ decay a singlet and its quark endpoint a triplet therefore describes the combined charge and the constituent charge, respectively. The endpoint colours are correlated to give a colourless pair.

\begin{figure}[htbp]
\centering
\begin{minipage}[t]{.32\linewidth}\centering

\begin{tikzpicture}[x=.70cm,y=.74cm,font=\footnotesize,>=Stealth,
 fermion/.style={thick,postaction={decorate},decoration={markings,mark=at position .55 with {\arrow{Stealth}}}},
 boson/.style={thick,decorate,decoration={snake,amplitude=1.2pt,segment length=5pt}},
 gluon/.style={thick,decorate,decoration={coil,aspect=.7,amplitude=2pt,segment length=4pt}},
 stringA/.style={very thick,dashed,blue!75!black},
 stringB/.style={very thick,dashed,orange!85!black}]
\path[use as bounding box] (-.3,-4.05) rectangle (5.6,3.35);

\node[font=\footnotesize] at (2.65,3.1) {(a) $Z\to u_1\bar u_2s_3\bar s_4$};
\coordinate (v) at (1,0);
\coordinate (e) at (1.85,1.05);
\coordinate (g) at (2.8,0);
\coordinate (q1) at (4,2.1);
\coordinate (q4) at (4,.75);
\coordinate (q3) at (4,-.75);
\coordinate (q2) at (4,-2.1);
\draw[boson] (0,0)--node[above=4pt] {$Z$} (v);
\draw[fermion] (v)--(e);
\draw[fermion] (e)--(q1);
\draw[fermion] (q2)--(v);
\draw[gluon] (e)--node[above right=2pt] {$g^*$} (g);
\draw[fermion] (q4)--(g);
\draw[fermion] (g)--(q3);
\foreach \v in {v,e,g} \fill (\v) circle (1.5pt);
\node[above left=1pt] at (q1) {$u_1^a$};
\node[above left=1pt] at (q4) {$\bar s_4^{\bar a}$};
\node[below left=1pt] at (q3) {$s_3^b$};
\node[below left=1pt] at (q2) {$\bar u_2^{\bar b}$};
\draw[stringA] (q1) to[out=0,in=0,looseness=1.7] node[right,text=blue!75!black] {$(14)$} (q4);
\draw[stringB] (q3) to[out=0,in=0,looseness=1.7] node[right,text=orange!85!black] {$(32)$} (q2);
\node[align=center] at (2.65,-3.55) {$\mathcal A:\ \mathcal O(\alpha_s)$\\$\sigma:\ \mathcal O(\alpha_s^2)$};
\end{tikzpicture}
\end{minipage}
\hfill
\begin{minipage}[t]{.32\linewidth}\centering

\begin{tikzpicture}[x=.70cm,y=.74cm,font=\footnotesize,>=Stealth,
 fermion/.style={thick,postaction={decorate},decoration={markings,mark=at position .55 with {\arrow{Stealth}}}},
 boson/.style={thick,decorate,decoration={snake,amplitude=1.2pt,segment length=5pt}},
 gluon/.style={thick,decorate,decoration={coil,aspect=.7,amplitude=2pt,segment length=4pt}},
 stringA/.style={very thick,dashed,blue!75!black},
 stringB/.style={very thick,dashed,orange!85!black}]
\path[use as bounding box] (-.3,-4.05) rectangle (5.6,3.35);
\node at (2.65,3.1) {(b) $Z\to u_1\bar u_2g_3g_4$};
\coordinate (v) at (1,0);
\coordinate (e1) at (2.6,-1.3);
\coordinate (e2) at (2.6,1.3);
\coordinate (q1) at (4,2.1);
\coordinate (q2) at (4,-2.1);
\coordinate (g3) at (4,.75);
\coordinate (g4) at (4,-.75);
\draw[boson] (0,0)--node[above=4pt] {$Z$} (v);
\draw[fermion] (v)--(e2);
\draw[fermion] (e2)--(q1);
\draw[fermion] (q2)--(e1);
\draw[fermion] (e1)--(v);
\draw[gluon] (e2)--(g3);
\draw[gluon] (e1)--(g4);
\foreach \v in {v,e1,e2} \fill (\v) circle (1.5pt);
\node[above left=1pt] at (q1) {$u_1$};
\node[below left=1pt] at (q2) {$\bar u_2$};
\node[above right=2pt] at (g3) {$g_3$};
\node[below right=2pt] at (g4) {$g_4$};
\draw[stringA] (q1) .. controls (6.2,2.1) and (6.2,-2.1) ..
 node[pos=.28,right=2pt,text=blue!75!black] {$(12)$} (q2);
\draw[stringB] (g3) to[out=0,in=0,looseness=1.8] (g4);
\node[text=orange!85!black] at (4.1,0) {$(34)$};
\draw[stringB] (g4) to[out=180,in=180,looseness=1.8] (g3);
\node[align=center] at (2.65,-3.55) {$\mathcal A:\ \mathcal O(\alpha_s)$\\$\sigma:\ \mathcal O(\alpha_s^2)$};
\end{tikzpicture}
\end{minipage}
\hfill
\begin{minipage}[t]{.32\linewidth}\centering

\begin{tikzpicture}[x=.70cm,y=.74cm,font=\footnotesize,>=Stealth,
 fermion/.style={thick,postaction={decorate},decoration={markings,mark=at position .55 with {\arrow{Stealth}}}},
 boson/.style={thick,decorate,decoration={snake,amplitude=1.2pt,segment length=5pt}},
 gluon/.style={thick,decorate,decoration={coil,aspect=.7,amplitude=2pt,segment length=4pt}},
 stringA/.style={very thick,dashed,blue!75!black},
 stringB/.style={very thick,dashed,orange!85!black}]
\path[use as bounding box] (-.3,-4.05) rectangle (5.6,3.35);

\node[font=\footnotesize] at (2.65,3.1) {(c) $W^+W^-\to4q$};
\coordinate (vp) at (1.5,1.4);
\coordinate (vm) at (1.5,-1.4);
\coordinate (q1) at (4,2.1);
\coordinate (q2) at (4,.75);
\coordinate (q3) at (4,-.75);
\coordinate (q4) at (4,-2.1);
\draw[boson] (0,1.4)--node[above=4pt] {$W^+$} (vp);
\draw[boson] (0,-1.4)--node[above=4pt] {$W^-$} (vm);
\draw[fermion] (vp)--(q1);
\draw[fermion] (q2)--(vp);
\draw[fermion] (vm)--(q3);
\draw[fermion] (q4)--(vm);
\fill (vp) circle (1.5pt); \fill (vm) circle (1.5pt);
\node[above left=1pt] at (q1) {$u_1^a$};
\node[above left=1pt] at (q2) {$\bar d_2^{\bar a}$};
\node[below left=1pt] at (q3) {$s_3^b$};
\node[below left=1pt] at (q4) {$\bar c_4^{\bar b}$};
\draw[stringA] (q1) to[out=0,in=0,looseness=1.7] node[right,text=blue!75!black] {$(12)$} (q2);
\draw[stringB] (q3) to[out=0,in=0,looseness=1.7] node[right,text=orange!85!black] {$(34)$} (q4);
\node[align=center] at (2.65,-3.55) {$\mathcal A:\ \mathcal O(\alpha_s^0)$\\$\sigma:\ \mathcal O(\alpha_s^0)$};
\end{tikzpicture}
\end{minipage}
\caption{Three colour configurations. (a) Gluon splitting gives leading-colour strings $(14)(32)$, while the direct singlet pair $(12)(34)$ has zero projection in the exact hard state. (b) Two gluon emissions admit the separate $u_1\bar u_2$ and $g_3g_4$ singlets shown, as an alternative to the leading-colour chain. (c) Independent $W$ decays initially produce singlets $(12)(34)$.}
\label{fig:fourquarkorigin}
\end{figure}
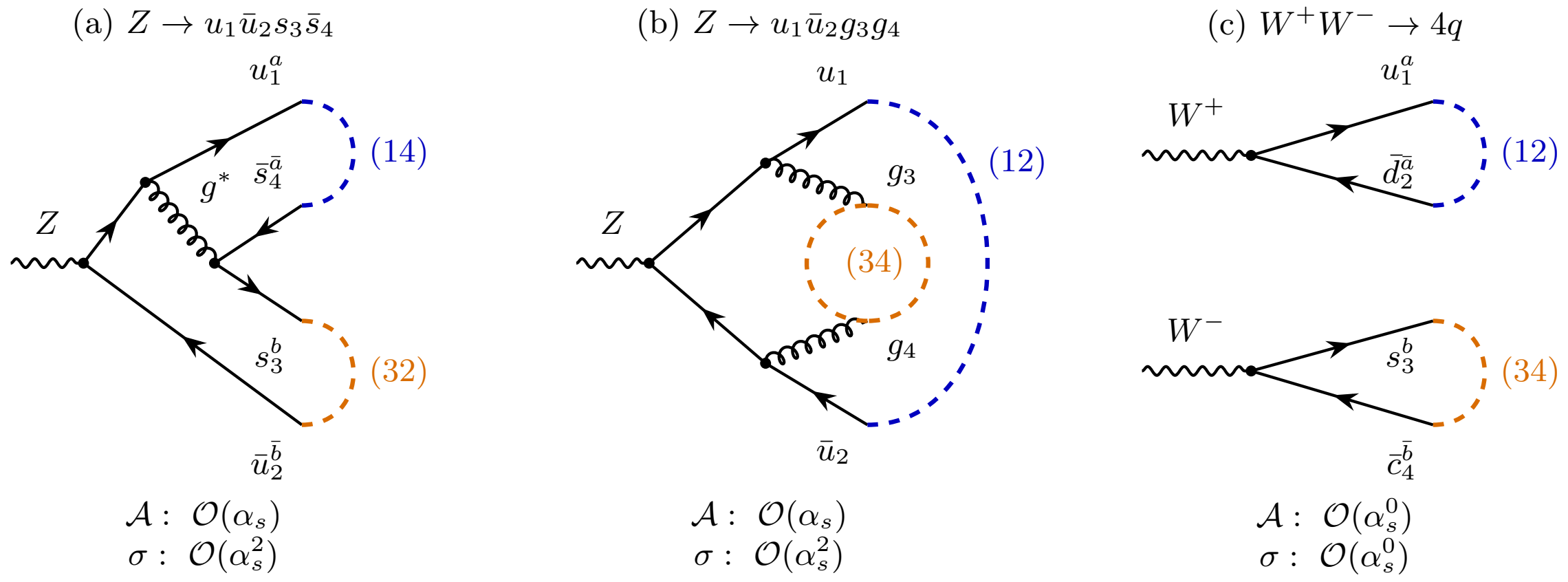

Leading colour means retaining the dominant terms as the number of colours $N_c$ becomes large, dropping contributions suppressed by inverse powers of $N_c$~\cite{tHooft:1973alw,Platzer:2013fha}. A gluon is represented by a colour and an anticolour, and matching colour lines define the string connections. One emission gives the chain $u_1-g-\bar u_2$. With electroweak factors suppressed and $\alpha_s=g_s^2/(4\pi)$, its single strong vertex gives an amplitude starting at order $g_s\sim\sqrt{\alpha_s}$ and a cross section starting at order $\alpha_s$ for the full $e^+e^-$ process.

Now let the emitted gluon split, $g^*\to s_3\bar s_4$, as in Fig.~\ref{fig:fourquarkorigin}(a). The leading-colour chain becomes the two strings $(14)(32)$. The daughter pair inherits the parent's octet charge. Splitting adds a second strong vertex, so the amplitude starts at order $g_s^2\sim\alpha_s$ and the cross section at order $\alpha_s^2$~\cite{Ballestrero:1996tu}. For this gluon-splitting contribution, the colour factor of the four-quark amplitude is~\cite{Ballestrero:1996tu,Platzer:2013fha}
\begin{equation}
 \mathcal A_{\mathrm{colour}}\propto
 \underbrace{\delta_{il}\delta_{kj}}_{(14)(32)}
 -\frac1{N_c}\underbrace{\delta_{ij}\delta_{kl}}_{(12)(34)}.
\end{equation}
Here $(i,j,k,l)$ are the colour indices of legs $(1,2,3,4)$, with $j,l$ labelling anticolours. Each Kronecker delta pairs a quark colour with an antiquark anticolour. The second pairing enters with a coefficient suppressed by $1/N_c$ and a relative minus sign.

Write $A$ for the normalized state of the direct singlets $(12)(34)$ and $B$ for that of the crossed singlets $(14)(32)$. The overlap amplitude between the normalized pairings is $1/N_c$, so projecting onto the direct singlets cancels the two contributions exactly. Destructive interference removes the component in which $(12)$ and $(34)$ are both singlets, leaving two octets coupled to an overall singlet. Keeping only $B$ assigns probability $1/9$ to the direct singlet pair at $N_c=3$, where the exact amplitude gives zero.

A reconnection model retaining these hard-state correlations must reject $(12)(34)$ as a probabilistic rearrangement of the two drawn strings. A kinematic criterion could favour that pairing if it shortened the strings, and would then select a channel with zero hard-state projection. Colour compatibility must therefore be checked before ranking candidates by string length. Amplitude-level soft-gluon evolution provides a dynamical approach to subsequent colour rearrangement, including baryonic channels~\cite{Gieseke:2018gff}.

Return to the original $Z\to u_1\bar u_2$ system and now emit two gluons. In $Z\to u_1\bar u_2g_3g_4$, the leading-colour chain can be $u_1-g_3-g_4-\bar u_2$. The two gluons can also carry compensating colour and form a singlet, leaving $u_1\bar u_2$ in a separate singlet. Figure~\ref{fig:fourquarkorigin}(b) shows this allowed rearrangement as an open quark string and a closed two-gluon string. Such separate gluonic singlets and their observable consequences were studied in Refs.~\cite{Friberg:1996xc,Wang:1999xz}. This channel is suppressed at large $N_c$~\cite{Wang:2000bv,Platzer:2013fha,ALEPH:2006jcu}. A colour-faithful reconnection model should include it among the possibilities whose realization depends on dynamics and space-time overlap. The two emission vertices give an amplitude starting at order $\alpha_s$ and a cross section starting at order $\alpha_s^2$, as for the four-quark history in panel (a).

Hadronic $WW$ decays provide two independent singlets directly. For example, $W^+\to u_1\bar d_2$ and $W^-\to s_3\bar c_4$ initially give the strings $(12)(34)$ in Fig.~\ref{fig:fourquarkorigin}(c), corresponding to the colour state $A$~\cite{Sjostrand:1993hi}. Both electroweak decays occur at strong-coupling order $\alpha_s^0$. The two singlets are thus available already at Born level, avoiding the $\alpha_s^2$ production suppression of the two-gluon $Z$ example. This makes hadronic $WW$ decays a much better suited system for studying colour-reconnection effects.

\subsection{From two strings to junctions and ropes}
\label{sec:fullrope}
Focus on the two $W$ decays and imagine that the quark ends $1$ and $3$ lie close together, with the antiquark ends farther away. A field that does not resolve the separation of the two quarks responds to their combined colour charge. We must therefore combine two triplet representations. Their nine colour states separate as
\begin{equation}\label{eq:qqrepresentations}
 3\otimes3=\bar3\oplus6.
\end{equation}
The antitriplet consists of three combinations antisymmetric in the two colour indices; the sextet consists of six symmetric combinations. Both are possible even though the complete four-parton event is a singlet. The antiquark pair carries the conjugate representation, so that either $\bar3\otimes3$ or $6\otimes\bar6$ can be coupled to total colour zero. We have regrouped the same four charges into a quark pair and an antiquark pair.

For independent singlet $W$ decays, tracing over the antiquarks leaves the two quarks uniformly distributed over their nine colour states. Their representation probabilities are therefore $P(\bar3)=3/9=1/3$ and $P(6)=6/9=2/3$, before dynamical selection of a string topology.

To see the antitriplet, choose red and blue for the two nearby quarks. A colour-singlet baryon combines red, blue and green antisymmetrically, so the red--blue pair must carry the colour conjugate to green:
\begin{equation}\label{eq:antigreenpair}
 |\bar g\rangle_{13}\equiv
 \frac{|r\rangle_1|b\rangle_3-|b\rangle_1|r\rangle_3}{\sqrt2}.
\end{equation}
This state transforms as antigreen, up to a phase convention; the other two antisymmetric pairs similarly transform as antired and antiblue~\cite{Christiansen:2015yqa,Sjostrand:2002ip}. At distances where the two constituents are unresolved, it acts as one anticolour charge and can be screened by green. A fixed red--blue product also contains a symmetric component: only its antisymmetric projection has this transformation law. The colour names specify a basis, while the multiplet distinction is basis independent.

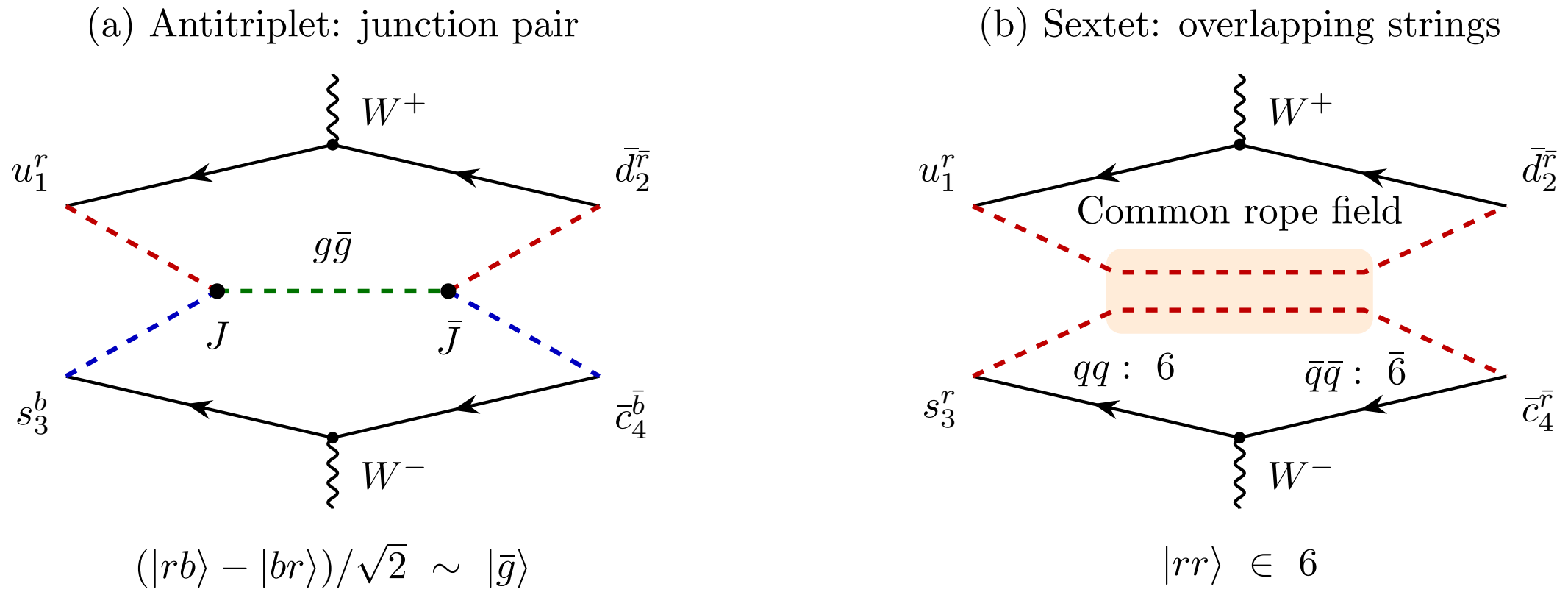
\begin{figure}[htbp]
\centering
\begin{minipage}[t]{.49\linewidth}\centering
\begin{tikzpicture}[x=.80cm,y=.85cm,font=\small,>=Stealth,
 fermion/.style={thick,postaction={decorate},decoration={markings,mark=at position .55 with {\arrow{Stealth}}}},
 boson/.style={thick,decorate,decoration={snake,amplitude=1.2pt,segment length=5pt}},
 string/.style={very thick,dashed}]
\path[use as bounding box] (-.7,-1.5) rectangle (6.7,4.7);
\node at (3,4.5) {(a) Antitriplet: junction pair};
\coordinate (vp) at (3,3.25); \coordinate (vm) at (3,.15);
\coordinate (q1) at (0,2.6); \coordinate (q2) at (6,2.6);
\coordinate (q3) at (0,.8); \coordinate (q4) at (6,.8);
\coordinate (j) at (1.7,1.7); \coordinate (aj) at (4.3,1.7);
\draw[boson] (3,4)--node[right=3pt] {$W^+$} (vp);
\draw[boson] (3,-.6)--node[right=3pt] {$W^-$} (vm);
\draw[fermion] (vp)--(q1); \draw[fermion] (q2)--(vp);
\draw[fermion] (vm)--(q3); \draw[fermion] (q4)--(vm);
\foreach \v in {vp,vm} \fill (\v) circle (1.5pt);
\node[above left] at (q1) {$u_1^r$}; \node[above right] at (q2) {$\bar d_2^{\bar r}$};
\node[below left] at (q3) {$s_3^b$}; \node[below right] at (q4) {$\bar c_4^{\bar b}$};
\draw[string,red!75!black] (q1)--(j); \draw[string,blue!75!black] (q3)--(j);
\draw[string,green!45!black] (j)--node[above=3pt,text=black] {$g\bar g$} (aj);
\draw[string,red!75!black] (aj)--(q2); \draw[string,blue!75!black] (aj)--(q4);
\fill (j) circle (2pt); \fill (aj) circle (2pt);
\node[below=4pt,align=center] at (j) {$J$};
\node[below=4pt,align=center] at (aj) {$\bar J$};
\node at (3,-1.2) {$(|rb\rangle-|br\rangle)/\sqrt2\ \sim\ |\bar g\rangle$};
\end{tikzpicture}
\end{minipage}\hfill
\begin{minipage}[t]{.49\linewidth}\centering
\begin{tikzpicture}[x=.80cm,y=.85cm,font=\small,>=Stealth,
 fermion/.style={thick,postaction={decorate},decoration={markings,mark=at position .55 with {\arrow{Stealth}}}},
 boson/.style={thick,decorate,decoration={snake,amplitude=1.2pt,segment length=5pt}},
 string/.style={very thick,dashed}]
\path[use as bounding box] (-.7,-1.5) rectangle (6.7,4.7);
\node at (3,4.5) {(b) Sextet: overlapping strings};
\coordinate (vp) at (3,3.25); \coordinate (vm) at (3,.15);
\coordinate (q1) at (0,2.6); \coordinate (q2) at (6,2.6);
\coordinate (q3) at (0,.8); \coordinate (q4) at (6,.8);
\draw[boson] (3,4)--node[right=3pt] {$W^+$} (vp);
\draw[boson] (3,-.6)--node[right=3pt] {$W^-$} (vm);
\draw[fermion] (vp)--(q1); \draw[fermion] (q2)--(vp);
\draw[fermion] (vm)--(q3); \draw[fermion] (q4)--(vm);
\foreach \v in {vp,vm} \fill (\v) circle (1.5pt);
\node[above left] at (q1) {$u_1^r$}; \node[above right] at (q2) {$\bar d_2^{\bar r}$};
\node[below left] at (q3) {$s_3^r$}; \node[below right] at (q4) {$\bar c_4^{\bar r}$};
\fill[orange!15,rounded corners=4pt] (1.5,1.25) rectangle (4.5,2.15);
\draw[string,red!75!black] (q1)--(1.6,1.9)--(4.4,1.9)--(q2);
\draw[string,red!75!black] (q3)--(1.6,1.5)--(4.4,1.5)--(q4);
\node[above] at (3,2.18) {Common rope field};
\node[below,align=center] at (1.7,1.2) {$qq:\ 6$};
\node[below,align=center] at (4.3,1.2) {$\bar q\bar q:\ \bar6$};
\node at (3,-1.2) {$|rr\rangle\ \in\ 6$};
\end{tikzpicture}
\end{minipage}
\caption{Representative colour components of two possible field configurations for nearby endpoints from the two $W$ strings. (a) The antisymmetric red--blue pair transforms as antigreen and permits a junction--antijunction connection, with a compensating green leg at the junction. (b) Two red quarks lie in the sextet; their overlapping strings can carry a common rope field, balanced by the antiquark antisextet.}
\label{fig:junctionrope}
\end{figure}

In a string description, the antitriplet coupling permits the two quark legs to meet at a junction. A third string leg carries the compensating colour, in place of a third external quark. At the opposite end, the two antiquark legs can meet at an antijunction, with the junction and antijunction joined by that internal string~\cite{Christiansen:2015yqa,Sjostrand:2002ip}. This gives the double-junction interpretation of the same four-parton system, illustrated in Fig.~\ref{fig:junctionrope}(a). The colour algebra establishes that this configuration is allowed. Its formation depends on the dynamics and geometry. Its fragmentation offers a route to baryon--antibaryon production while preserving the event's net baryon number. Pair creation on the internal string supplies a quark for the third junction leg and an antiquark for the third antijunction leg. For a surviving junction, fragmentation proceeds inwards along its legs, leaving three quarks to form a baryon at the junction and three antiquarks to form an antibaryon at the antijunction~\cite{Sjostrand:2002ip,Bierlich:2023okq}. This supplies baryon formation through the string topology, in addition to the ordinary diquark--antidiquark pair-production channel~\cite{Andersson:1984af}. A charm quark retained on a junction leg can thereby enter a charm baryon.

Now give both nearby quarks the same colour, red. The state $|r\rangle_1|r\rangle_3$ is symmetric and lies entirely in the sextet. There is no antisymmetric red--red combination, so these two charges cannot act as a single anticolour. The six symmetric states comprise the three equal-colour pairs and the three symmetric combinations of unequal colours. If the two fields overlap, they can support a common field carrying this sextet charge, balanced by an antisextet at the other end. This is the simplest example of a higher-representation rope, illustrated in Fig.~\ref{fig:junctionrope}(b)~\cite{Bierlich:2014xba,Bierlich:2017sxk}. Its field strength requires an additional physical assumption, developed below.

Junctions and ropes are thus two outcomes of the same colour decomposition. An antitriplet pair can join a fundamental string through a junction, while a sextet pair supports a higher-representation field. For opposite orientations, the $3\otimes\bar3=1\oplus8$ decomposition instead permits either colour cancellation or an octet field. 

\subsection{Screening and flavour production}
\label{sec:ropescreening}\label{sec:fragmap}
When strings overlap within a limited transverse region, their colour fields can act coherently and form a rope \cite{Biro:1984cf}. The combined endpoint charges then determine the colour representation carried by the common field. To describe its fragmentation, we need the energy stored per unit length: the rope tension. Lattice calculations of potentials between static colour sources support approximate scaling with the quadratic Casimir, which measures the squared colour charge of a representation~\cite{Bali:2000un}. Following the rope model, we use this scaling to assign tensions to overlapping fields and normalize them to the tension $\kappa$ of a fundamental string~\cite{Bierlich:2014xba}:
\begin{equation}\label{eq:casimir}
 \kappa_{p,q}=\kappa\frac{C_2(p,q)}{C_2(1,0)}
 =\frac\kappa4(p^2+pq+q^2+3p+3q).
\end{equation}
Here $(p,q)$ labels the representation: $3=(1,0)$, $\bar3=(0,1)$, $6=(2,0)$ and $8=(1,1)$. Thus a sextet has tension $5\kappa/2$, an octet $9\kappa/4$, and a triplet or antitriplet $\kappa$. Static-source calculations motivate Casimir scaling over an intermediate-distance range. Applying it to dynamically overlapping strings is a model assumption which goes beyond colour reconnection as introduced in the previous section.

\begin{figure}[htbp]
\centering
\begin{tikzpicture}[>=Stealth,font=\small,x=1cm,y=1cm,
 string/.style={very thick,dashed,postaction={decorate},decoration={markings,mark=at position .5 with {\arrow{Stealth}}}}]
\draw[string,blue!75!black] (.2,3.05)--(1.8,3.05);
\draw[string,orange!85!black] (.2,2.72)--(1.8,2.72);
\draw[string,blue!75!black] (7.2,3.05)--(8.8,3.05);
\draw[string,orange!85!black] (8.8,2.72)--(7.2,2.72);
\draw[string,blue!75!black] (2.2,1.72)--(3.8,1.72);
\draw[string,orange!85!black] (10.8,1.72)--(9.2,1.72);
\foreach \x/\y/\pq in {1/3.5/{2,0},3/2.2/{1,0},5/.65/{0,0},8/3.5/{1,1},10/2.2/{0,1},12/.65/{0,0}} {
 \node at (\x,\y) {$(p,q)=(\pq)$}; }
\foreach \x/\y/\lab in {1/2/{5\kappa/2},3/1/{\kappa},5/0/{0},8/2/{9\kappa/4},10/1/{\kappa},12/0/{0}} {
 \draw[thick] (\x-.7,\y)--(\x+.7,\y);
 \node[above] at (\x,\y) {$\lab$}; }
\draw[->,thick,orange!80!black] (1.5,1.9)--(2.5,1.05) node[midway,below left] {$F=3\kappa/2$};
\draw[->,thick] (3.5,.9)--(4.5,.05) node[midway,below left] {$F=\kappa$};
\draw[->,thick,orange!80!black] (8.5,1.9)--(9.5,1.05) node[midway,below left] {$F=5\kappa/4$};
\draw[->,thick] (10.5,.9)--(11.5,.05) node[midway,below left] {$F=\kappa$};
\end{tikzpicture}
\caption{Sequential rope screening for parallel strings (left) and oppositely oriented strings (right). Each rung gives the representation labels $(p,q)$ and its tension; transitions give the force available to each breakup.}
\label{fig:unified}
\end{figure}
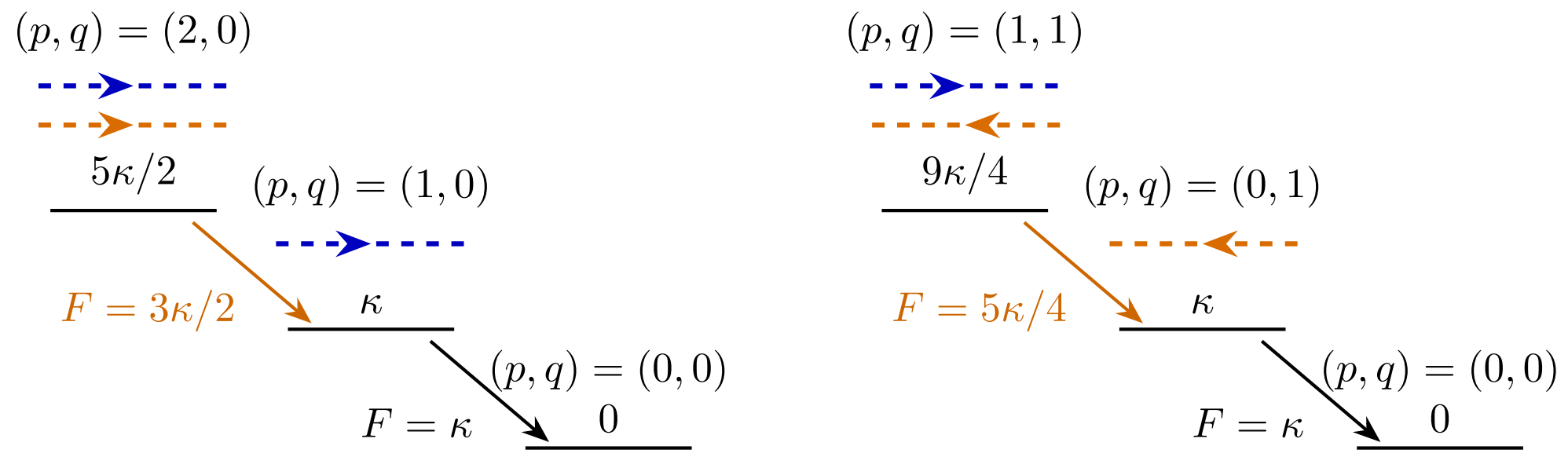

The energy released by screening determines the force governing string breaking. If a newly created pair separates by $\ell$ and changes the field between its members from $R$ to $R'$, the available energy is $(\kappa_R-\kappa_{R'})\ell$. As the name ``rope hadronization'' suggests, one should imagine individual strings joining together as strands in a rope. The rope then breaks one strand at a time. The force entering the tunnelling calculation of Section~\ref{sec:wkb} is therefore
\begin{equation}\label{eq:ropeforce}
 F=\kappa_R-\kappa_{R'},\qquad
 \rho_s(F)=\rho_0^{\kappa/F}.
\end{equation}
Figure~\ref{fig:unified} follows the screening of two parallel strings through the sextet channel (left) and two oppositely oriented strings through the octet channel (right). The red--red sextet follows $6\to3\to1$: the first step releases $5\kappa/2-\kappa=3\kappa/2$, and the remaining triplet breaks with force $\kappa$. The octet follows $8\to\bar3\to1$, releasing $5\kappa/4$ and then $\kappa$. In each ladder the successive releases sum to the initial tension. For the illustrative baseline $\rho_0=0.30$, the first strange weights are $0.448$ and $0.382$, respectively; both return to $0.30$ at the fundamental step. For independent, uniformly populated endpoint colours, the sextet and octet channels have probabilities $2/3$ and $8/9$, respectively, before conditioning by CR.

A sextet region supplies an enhanced screening step followed by an ordinary one, whereas an antitriplet carries one fundamental unit of field. Representation probabilities therefore cannot simply be identified with fractions of enhanced hadrons. The resulting hadron yields depend on how much string fragments at each force and on the quarks and diquarks produced in its breaks. Baryon production also depends on how the available momentum is shared and how junctions fragment~\cite{Bierlich:2017sxk,Bierlich:2022oja}. Measuring both baryon-to-meson ratios and the strange fraction within each hadron family helps distinguish changes in baryon formation from changes in strange-quark production.

\subsection{The experimental two-string test}\label{sec:wwtest}
The two hadronic $W$ decays suggest a direct experimental test: does a second hadronically decaying $W$ change the flavour composition? Fully hadronic $WW\to4q$ events can be compared with the hadronic side of semileptonic $WW\to2q\ell\nu$ events. Define the ratio of mean yields
\begin{equation}\label{eq:wwflavourdouble}
 D_h=\frac{(\langle N_h\rangle/\langle N_{\pi^\pm}\rangle)_{4q}}
 {(\langle N_h\rangle/\langle N_{\pi^\pm}\rangle)_{2q\ell\nu}}.
\end{equation}
For independent hadronization and matched acceptance and decay-flavour composition, $D_h=1$. Both samples have the same hadronic mass scale, $m_W$, and primary weak-decay flavour mixture.

DELPHI published the requisite charged-pion, kaon and proton multiplicities at $\sqrt{s}=189\,\mathrm{GeV}$~\cite{DELPHI:2000ahn}. We construct the double ratios in Table~\ref{tab:wwflavourdouble} from those published yields. Their uncertainties leave modest flavour enhancements unresolved.
\begin{table}[htbp]
\centering
\begin{tabular}{lccc}
\toprule
Species $h$ & $D_h$ & Statistical uncertainty & Systematic uncertainty\\
\midrule
$K^++K^-$ & 0.963 & 0.168 & 0.085\\
$p+\bar p$ & 0.949 & 0.278 & 0.108\\
\bottomrule
\end{tabular}
\caption{Flavour double ratios at $\sqrt{s}=189\,\mathrm{GeV}$, calculated from Table~5 of Ref.~\cite{DELPHI:2000ahn}. All four input yields are treated as uncorrelated, separately for statistical and systematic errors. The table supplies no covariance matrix; particle-identification and common systematic correlations could change these uncertainties. Adding the displayed errors in quadrature gives approximately 0.19 and 0.30.}
\label{tab:wwflavourdouble}
\end{table}

The missing measurements are matched $4q$ and $2q\ell\nu$ yields for $\Lambda$, $\Xi$ and $\Omega$, including antiparticles, together with their pion reference and yield covariances. These are absent from the DELPHI $WW$ tables; its $\Lambda$ results concern continuum $q\bar q$ production. A continuum sample cannot simply replace the single-$W$ reference. At the same collider energy, the continuum hadronic system has invariant mass near $\sqrt{s}$, whereas each $W$ system has mass near $m_W$; $Z$-pole data provide yet another scale. The primary quark-flavour mixtures also differ, affecting secondary hadrons from heavy-flavour decays. Transporting a continuum flavour ratio to a hadronic $W$ therefore introduces fragmentation and flavour-composition modelling uncertainties that the channel-matched comparison largely avoids.

Reanalyses of archived ALEPH and DELPHI data demonstrate the continuing scope for new electron--positron measurements~\cite{Electron-PositronAlliance:2021kig}. FCC-ee offers a much larger prospective $W$ sample and a dedicated hadronization and interconnection programme~\cite{Christiansen:2015yca,Ploerer:2022ftv}; its hyperon reach will depend on reconstruction and systematic uncertainties.

For now, pp collisions provide the statistically powerful, species-resolved evidence for multiplicity-dependent strangeness enhancement, including multistrange baryons~\cite{ALICE:2016fzo,ALICE:2019avo}. We therefore need the pp measurements to confront the proposed flavour mechanism where an effect is already resolved, and subsequently follow it into nuclear collisions. The greater density of possible string partners comes with additional geometry and production biases: an enhancement in pp alone cannot establish its microscopic origin.

\section{From toy systems to real collisions}\label{sec:collisions}
The colour calculation identifies allowed channels. Their realization depends on the momenta and space-time evolution of the fields. In $W^+W^-$ events, the $W$ lifetimes determine the decay vertices, and the evolving strings must overlap before fragmentation for reconnection to occur. Hadron and nuclear collisions introduce many possible partners at different positions. We review how implementations select them, then introduce local escape geometry and a moderated large-force response for the extension from pp to p--Pb and O--O.
\subsection{Kinematics and locality of string interactions}\label{sec:implementation}
Multiple parton interactions and beam remnants supply additional candidates in hadron collisions~\cite{Sjostrand:1987su,Bierlich:2022pfr}. Early dipole reconnection models implemented rearrangements in Ariadne and studied their consequences at LEP~\cite{Lonnblad:1995yk}. The QCD-inspired reconnection model combines finite-colour compatibility with a preference for smaller string length~\cite{Christiansen:2015yqa}.

For the two strings from the $W$ example, compare the original connections $(12)(34)$ with the alternative $(14)(32)$. For pair masses large compared with a reference scale $m_0$, a representative length measure and its change are
\begin{equation}\label{eq:lengthcriterion}
 \lambda\simeq\sum_{\text{adjacent }ij}\ln\frac{s_{ij}}{m_0^2},
 \qquad
 \Delta\lambda\equiv\lambda_{14,32}-\lambda_{12,34}
 \simeq\ln\frac{s_{14}s_{32}}{s_{12}s_{34}},
 \qquad s_{ij}=(p_i+p_j)^2.
\end{equation}
If, for example, both new pair masses are half the original pair masses, $\Delta\lambda=-\ln16$, and the alternative offers less longitudinal phase space for fragmentation. The model favours such reductions among colour-compatible candidates, subject also to its causality constraints. The low-mass implementation is regulated; $m_0$ here is a reconnection reference scale. Junction configurations require a corresponding measure for their legs.

A heavy endpoint shows why that momentum-space measure needs care. For a light junction leg, a common prescription uses its energy $E$ in the junction rest frame. For a heavy-quark leg, Ref.~\cite{Bierlich:2023okq} instead uses the quark's rapidity in that frame:
\begin{equation}\label{eq:junctionmassmeasure}
 \lambda_j=\ln\!\left(1+\frac{\sqrt2 E}{m_j}\right),
 \qquad
 \lambda_{\rm HQ}=\frac12\ln\frac{E+p}{E-p}
 =\operatorname{arsinh}\frac{p}{m}.
\end{equation}
Here $p$ is the momentum magnitude, $m$ the heavy-quark mass and $m_j$ a tunable reference scale for the light-leg measure. As $p\to0$, the heavy quark retains its rest energy, so the energy-based expression remains finite while its rapidity vanishes. Rest mass does not provide a long rapidity interval for string fragmentation. Correcting this distinction changes the ranking of junction configurations involving heavy quarks and hence their opportunity to form charm baryons. Together with improved junction fragmentation, this was developed for the Angantyr description of charm production in pp and p--Pb~\cite{Bierlich:2023okq}.

Nuclear collisions additionally require spatial information. Translating one of the two strings several femtometres changes none of the invariants in Eq.~\eqref{eq:lengthcriterion}; a favourable $\Delta\lambda$ therefore cannot establish that the strings can communicate. This becomes essential when candidate partners originate in different nucleon sub-collisions. Spatially constrained reconnection allows nearby dipoles to reconnect across those sub-collisions while excluding distant candidates~\cite{Lonnblad:2023stc,Bierlich:2023okq}. The present calculation retains reconnection within nucleon sub-collisions, with junctions enabled; the additional global, spatially constrained reconnection stage is disabled. Rope overlap supplies the field interactions across sub-collisions.

Ropes face a related question: which fields coexist long enough to contribute to a common representation? Initially nearby pieces can separate before fragmentation. At fixed transverse momentum, its larger transverse mass gives a smaller transverse velocity. Below we use a simplified displacement estimate to decide whether a piece remains in the surrounding field. We first examine the colour correlations of that background, which geometrical overlap alone does not determine.

\subsection{Local overlap and the colour of the background}\label{sec:geometry}
A rope requires transverse overlap, a common longitudinal interval and compatible evolution times. The Lund space-time construction and the pairwise parallel-frame treatment provide a framework for evaluating these conditions~\cite{Ferreres-Sole:2018vgo,Bierlich:2020naj,Bierlich:2022oja}. The Gaussian field profiles and their shoving interaction are reviewed in Ref.~\cite{Bierlich:2024odg}; a complementary momentum-space construction describes the fragmentation of two repelling strings~\cite{Duncan:2019poz}. Shoving can dilute the string density before hadronization and thereby moderate the growth of rope tension~\cite{Chakraborty:2022nui}. The strings contributing to a breakup can themselves interact with a surrounding background. Geometrical overlap counts do not fully specify that background's correlated colour state.

Consider three strings $A,B,C$, with appreciable $AB$ and $BC$ overlaps but negligible direct $AC$ overlap, as sketched in Fig.~\ref{fig:threestringscreening}. The nearby ends are triplets, with conjugate ends completing the colour-singlet event. Suppose that $AB$ initially occupies a sextet. For fixed transverse width $R$ and separation $d_\perp$, normalize its interaction energy per unit common length to the string shoving expression~\cite{Bierlich:2024odg},
\begin{equation}\label{eq:backgroundoverlapenergy}
 E_{\rm int}(d_\perp)=2g\kappa\exp\!\left(-\frac{d_\perp^2}{4R^2}\right),
 \qquad
 E_{\rm int}^{(6)}=E_{\rm int},\quad
 E_{\rm int}^{(\bar3)}=-2E_{\rm int}.
\end{equation}
Here $g$ sets the overlap-energy normalization.\footnote{Matching the coincident overlap energies to the Casimir tensions gives $g=1/4$ in this toy. The relative reduction in Eq.~\eqref{eq:backgroundscreening} is independent of this normalization.} The relative interaction energies follow from the excess over two separated strings: $\kappa_6-2\kappa=\kappa/2$, whereas $\kappa_{\bar3}-2\kappa=-\kappa$. The negative interaction term denotes a reduction from the separated-string energy. Retaining both signs extends the usual repulsive shoving description, in which colour-reducing configurations are assumed to have been treated already~\cite{Bierlich:2020naj,Bierlich:2024odg}.

We model selection of the $BC$ antitriplet by projecting the initial colour state onto that channel and normalizing the resulting state. We then ask which $AB$ representation it contains. The same triplet $B$ participates in both couplings, so a definite $BC$ channel need not preserve a definite $AB$ channel. Take $AB$ initially in the symmetric sextet channel of $3_A\otimes3_B$, with $C$ independent and uniformly distributed over its three colours. The following conditional probabilities hold for any initial colour state within that sextet. SU(3) recoupling gives
\begin{equation}\label{eq:backgroundrecoupling}
 P(6_{AB}\mid\bar3_{BC})=\frac34,
 \qquad P(\bar3_{AB}\mid\bar3_{BC})=\frac14.
\end{equation}
Consequently, at unchanged $AB$ separation and profile,
\begin{equation}\label{eq:backgroundscreening}
 \left\langle E_{AB,\rm after}\right\rangle
 =\frac34 E_{\rm int}(d_{AB})
  +\frac14[-2E_{\rm int}(d_{AB})]
 =\frac14 E_{AB,\rm before}.
\end{equation}
An interaction with $C$ can therefore reduce the enhanced $AB$ overlap energy without a direct $AC$ contribution. What changes is the shared colour correlation; $B$'s quark endpoint remains a triplet.

\begin{figure}[htbp]
\centering
\begin{tikzpicture}[x=1.15cm,y=.95cm,font=\small,>=Stealth,
 fermion/.style={thick,postaction={decorate},decoration={markings,mark=at position .55 with {\arrow{Stealth}}}},
 boson/.style={thick,decorate,decoration={snake,amplitude=1.2pt,segment length=5pt}},
 string/.style={very thick,dashed}]
\path[use as bounding box] (-.65,-1.65) rectangle (10.55,2.65);
\foreach \x/\lab in {0/A,3/B,6/C} {
 \coordinate (q\lab) at (\x,2);
 \coordinate (a\lab) at (\x+3.8,.8);
 \path (q\lab) to[out=-10,in=90] coordinate[pos=.5] (mid\lab) (a\lab);
}
\begin{scope}[shift={($(midA)!.5!(midB)$)},rotate=-8]
 \filldraw[fill=violet!20,fill opacity=.7,draw=violet!55]
  (0,0) ellipse [x radius=1.55cm,y radius=.65cm];
\end{scope}
\begin{scope}[shift={($(midB)!.5!(midC)$)},rotate=8]
 \filldraw[fill=brown!22,fill opacity=.7,draw=brown!65]
  (0,0) ellipse [x radius=1.55cm,y radius=.65cm];
\end{scope}
\foreach \x/\lab/\col in {0/A/blue!75!black,3/B/red!80!black,6/C/green!45!black} {
 \coordinate (v\lab) at (\x+1.9,.2);
 \coordinate (q\lab) at (\x,2);
 \coordinate (a\lab) at (\x+3.8,.8);
 \draw[boson] (\x+1.9,-.65)--node[right=3pt] {$V_{\lab}$} (v\lab);
 \draw[fermion] (v\lab)--node[pos=.55,above right=2pt] {$q_{\lab}$} (q\lab);
 \draw[fermion] (a\lab)--node[pos=.55,below=3pt] {$\bar q_{\lab}$} (v\lab);
 \fill (v\lab) circle (1.5pt);
 \draw[string,\col] (q\lab) to[out=-10,in=90] (a\lab);
}
\node[font=\footnotesize] at ($ (midA)!.5!(midB)+(0,.25) $) {$AB:\ 3\otimes3\to6$};
\node[font=\footnotesize] at ($ (midB)!.5!(midC)+(0,.25) $) {$BC:\ 3\otimes3\to\bar3$};
\node[align=center] at (4.8,-1.25)
 {$AB:\quad 6\ \longrightarrow\ \tfrac34\,[6]+\tfrac14\,[\bar3]$\quad (conditional channel populations)};
\end{tikzpicture}
\caption{Three-string extension of the two-singlet example. Separate purple and brown shaded regions mark the $AB$ and $BC$ interactions, respectively. An initial $AB$ sextet, followed by selection of a $BC$ antitriplet, yields the conditional populations in Eq.~\eqref{eq:backgroundrecoupling}.}
\label{fig:threestringscreening}
\end{figure}

This conditional example shows how background colour correlations can moderate an enhanced overlap energy. Starting instead with three independent random colours, selecting the $BC$ antitriplet leaves the $AB$ sextet probability at $2/3$. A net screening effect requires dynamics that populate correlated configurations, with the other outcomes included. In separated overlap regions, their influence must also propagate within the available time. The existing PYTHIA implementation of rope hadronization does not take this kind of screening-by-background into account.

\subsection{Extension to nuclear collisions: geometry and saturation}\label{sec:nuclearimplementation}
Angantyr is the heavy ion extension to PYTHIA, and supplies explicit partonic systems and their nucleon-collision geometry~\cite{Bierlich:2018xfw,Miller:2007ri,Bierlich:2016smv}. Its wounded-nucleon construction respects the shared projectile and target energy~\cite{Bialas:1976ed,Andersson:1986gw}. Similar charged multiplicities can arise from a compact pp configuration or a more extended nuclear system, with different path lengths and overlap histories. Published cross-system rope calculations produce strangeness enhancement but also expose excessive baryon production in dense systems~\cite{Bierlich:2022ned}.

The extension separates two questions: 1) which moving pieces remain in the common field, and 2) how the correlated background affects its fragmentation response. For the first, each candidate string centre is surrounded by a transverse disc of radius $r_0$. Along the candidate's direction of motion, we merge the intervals covered by discs whose strings overlap its midpoint rapidity. The forward extent of the connected interval containing the candidate defines a path $L$. For a vanishing transverse-velocity proxy, the implementation imposes no escape restriction; the momentum ceiling still applies. All candidates enter this geometric estimate before the eligibility cut, avoiding dependence on their processing order.

The displacement estimate uses an average of endpoint transverse-velocity proxies,
\begin{equation}\label{eq:escapevelocity}
 \boldsymbol\beta_\perp=\frac12\left(\frac{\boldsymbol p_{\perp,1}}{m_{\perp,1}}
                  +\frac{\boldsymbol p_{\perp,2}}{m_{\perp,2}}\right),
 \qquad m_{\perp,i}=\sqrt{m_i^2+p_{\perp,i}^2}.
\end{equation}
The exact endpoint transverse velocity in a common frame is $\boldsymbol p_\perp/E$. The use of $m_\perp$ instead makes Eq.~\eqref{eq:escapevelocity} a proxy for transverse motion. The additional test uses the transverse momenta and string positions in the frame of the hadronizing event, with the beam defining the longitudinal axis. This estimate simplifies the parallel-frame construction~\cite{Bierlich:2020naj,Bierlich:2022oja}, where pairwise endpoint motion, relative orientation and proper-time-dependent field widths determine the evolving overlap. We retain the underlying pairwise rope treatment, but use one velocity proxy and a connected path for the additional eligibility test. Orientation changes and cancellations between endpoint velocities are treated only approximately.

The eligibility thresholds apply to the total dipole transverse momentum $|\boldsymbol p_{\perp,1}+\boldsymbol p_{\perp,2}|$, whereas the displacement uses the velocity proxy in Eq.~\eqref{eq:escapevelocity}. Every dipole below the existing soft threshold remains eligible. Additional dipoles up to a safety ceiling enter when $t_{\rm esc}|\boldsymbol\beta_\perp|<L$. The thresholds and propagation interval are specified in the reproducibility material. A massive endpoint moves more slowly at fixed transverse momentum, while an extended connected nuclear geometry offers a longer escape path. This local propagation criterion replaces a system-wide increase of the momentum cutoff. The effective propagation interval $t_{\rm esc}$ is a model parameter defined in the frame used for this estimate.

For the second extension, we represent the unresolved background response of Section~\ref{sec:geometry} by moderating the large-force tail after representation selection. For a raw screening-force ratio $h=F/\kappa$, define
\begin{equation}\label{eq:saturation}
 h_{\rm eff}=\begin{cases}
 h,&h\le h_0,\\
 h_0+(h_{\rm sat}-h_0)\left[1-e^{-(h-h_0)/(h_{\rm sat}-h_0)}\right],&h>h_0,
 \end{cases}\qquad h_{\rm sat}>h_0.
\end{equation}
The response joins the unsaturated sector with unit slope and approaches $h_{\rm sat}$ at large $h$. The working values $h_0=2$, $h_{\rm sat}=3$ preserve the fundamental, octet and sextet screening examples. The hypothesis is that collective background screening limits the coherent force available to an individual breakup, while additional strings can still increase the total energy and number of breakups. The limiting value and interpolation are phenomenological choices to be tested independently. Production also applies the ordinary-string lower bound, $h_{\rm frag}=\max\{1,h_{\rm eff}\}$, and uses $F_{\rm frag}=\kappa h_{\rm frag}$ in fragmentation.

Hadronic interactions continue after hadronization. Microscopic transport models such as UrQMD have long described this stage through hadron scattering and resonance formation and decay~\cite{Bleicher:1999xi}. Elastic collisions redistribute momenta and can change yields within a finite acceptance. Inelastic reactions, baryon--antibaryon annihilation and regeneration can change the particle abundances themselves. In central AA simulations, primary hadrons can undergo several subsequent collisions~\cite{daSilva:2020cyn,Bierlich:2021poz}.

EPOS provides a useful comparison because its core--corona construction can be followed by a hadronic cascade. Werner separates core, corona and full EPOS4 results, showing in particular how the afterburner modifies integrated baryon-to-pion ratios through annihilation at large multiplicity~\cite{Werner:2023jps}. EPOS3 calculations with and without UrQMD also demonstrate changes in resonance signals in Pb--Pb and p--Pb~\cite{Knospe:2015nva,Knospe:2021jgt}. Scattering of resonance decay daughters can remove a reconstructible signal even when those daughters survive as stable hadrons; regeneration can replenish the resonance population.

We stop at O--O to limit the subsequent evolution omitted from the present calculation. A quantitative Pb--Pb extension requires a hadronic stage; a rescattering framework is available within PYTHIA~\cite{Sjostrand:2020gyg} and is functional for nuclear collisions \cite{Bierlich:2021poz}. Rope formation can alter the hadron production vertices supplied to the cascade. Extending the present calculation to large nuclei requires a consistent treatment of that coupling.

\subsection{Comparison conventions}
The following comparisons use PYTHIA 8.319 with the combined colour reconnection beyond leading colour (CR-BLC) and rope prescription (solid curves) and the Monash baseline (dashed curves), with zero transverse shoving force and without hadronic rescattering. The Monash reference retains its ordinary reconnection model; the solid curves add the QCD-inspired junction prescription and rope treatment, together with their associated tune choices.

The event activity measures follows ALICE convention, reproducing the experimental setup, with analyses implemented in Rivet \cite{Bierlich:2024vqo}. The ALICE forward-detector acceptances are V0A ($2.8<\eta<5.1$) and V0C ($-3.7<\eta<-1.7$), with V0M combining their charged-particle activity. V0M selections define pp and O--O activity classes, while V0A defines p--Pb classes. Each model has its own frozen activity calibration. Charge-conjugate yields are summed.

For the activity-dependent curves, the generator-level V0AND trigger condition requires charged-particle activity on both forward sides; pp additionally requires at least one stable charged particle at $|\eta|<1$ (INEL$>0$). The horizontal coordinate is the class mean $\langle N_{\rm ch}(|\eta|<0.5)\rangle/\Delta\eta$, with $\Delta\eta=1$, primary charged particles and no transverse-momentum cut.

\subsection{Flavour ratios from pp to nuclear collisions}
In pp collisions, the rise of $K/\pi$, $\Lambda/\pi$, $\Xi/\pi$ and $\Omega/\pi$ with activity becomes stronger with increasing strangeness content~\cite{ALICE:2016fzo,ALICE:2019avo}. Strange-particle measurements therefore constrain aspects of hadronization left weakly determined by inclusive charged multiplicities~\cite{CMS:2011jlm,Skands:2014pea}.

Ropes can generate this hierarchy through the force-dependent flavour weights of Section~\ref{sec:ropescreening}, sampled over the breakup environments selected by colour and geometry~\cite{Bierlich:2014xba,Bierlich:2022oja}. Reconnection also changes string lengths, baryon formation and pion production. Tuning to the total charged multiplicity strongly constrains the pion yield, leaving its residual variation as a secondary contribution to the flavour ratios. The baryon-to-meson ratio $\Lambda/(2K_S^0)$ helps constrain these accompanying changes~\cite{Bierlich:2015rha}. 

Other descriptions of integrated yields include canonical conservation, core--corona evolution and thermodynamical fragmentation~\cite{Vovchenko:2019kes,Kanakubo:2019ogh,Pierog:2013ria,Fischer:2016zzs}. String closepacking also enhances strangeness through an increased effective tension, similar to ropes, and improves the description of the accompanying proton yield~\cite{Altmann:2025afh}. A time-dependent tension offers another source of enhanced flavour weights, with proposed tests in archived electron--positron data~\cite{Hunt-Smith:2020lul}.

\begin{figure}[htbp]
\centering
\includegraphics[width=\linewidth]{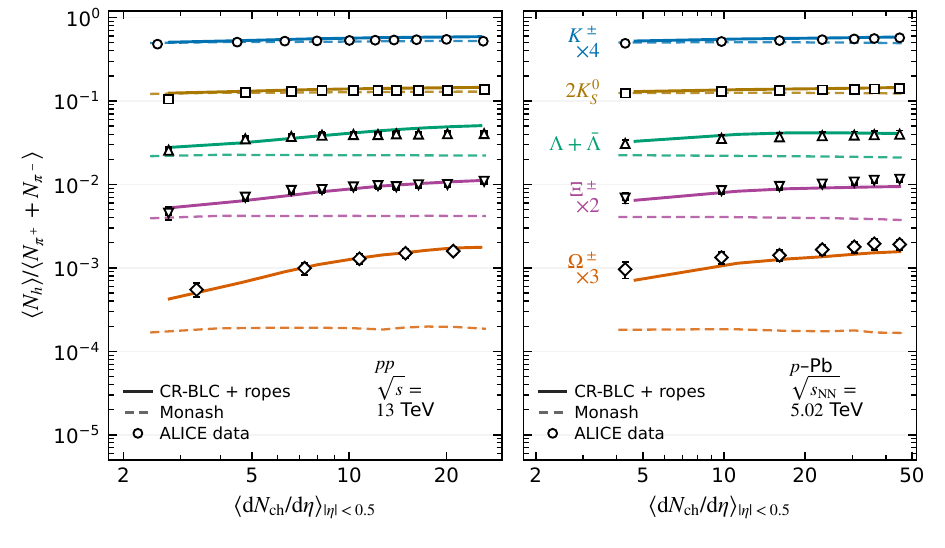}
\caption{Integrated strange-hadron/pion ratios versus charged-particle density in pp at 13 TeV (left) and p--Pb at 5.02 TeV (right). Open symbols are ALICE measurements~\cite{ALICE:2020nkc,ALICE:2019avo,ALICE:2013wgn,ALICE:2015mpp}; curves follow the conventions in the text. Primary light hadrons are selected at $|y|<0.5$ in pp and $-0.5<y_{\rm CM}<0$ in p--Pb, with the proton moving towards positive rapidity.}
\label{fig:pp}
\end{figure}

Figure~\ref{fig:pp} shows that the combined prescription, including the modifications described in Section~\ref{sec:nuclearimplementation}, gives a satisfactory description of the strange-hadron/pion ratios in pp and p--Pb. It reproduces the increasing hierarchy with strangeness content, with a much stronger activity response for $\Omega$ than for kaons.

Charm ratios probe how an existing heavy quark combines with the surrounding light flavours. ALICE fragmentation fractions show enhanced charm-baryon production in hadronic collisions relative to electron--positron measurements~\cite{ALICE:2020wfu,ALICE:2023sgl,ALICE:2024ocs}. Junction formation, additional excited baryons feeding down to $\Lambda_c$, and coalescence with surrounding quarks can enhance $\Lambda_c/D^0$~\cite{Bierlich:2023okq,He:2019tik,Minissale:2020bif}. Combining $\Lambda_c/D^0$ with $\Xi_c/\Lambda_c$ and $D_s/D^+$ tests both baryon formation and the strange fraction within each hadron family.

\begin{figure}[htbp]
\centering
\includegraphics[width=\linewidth]{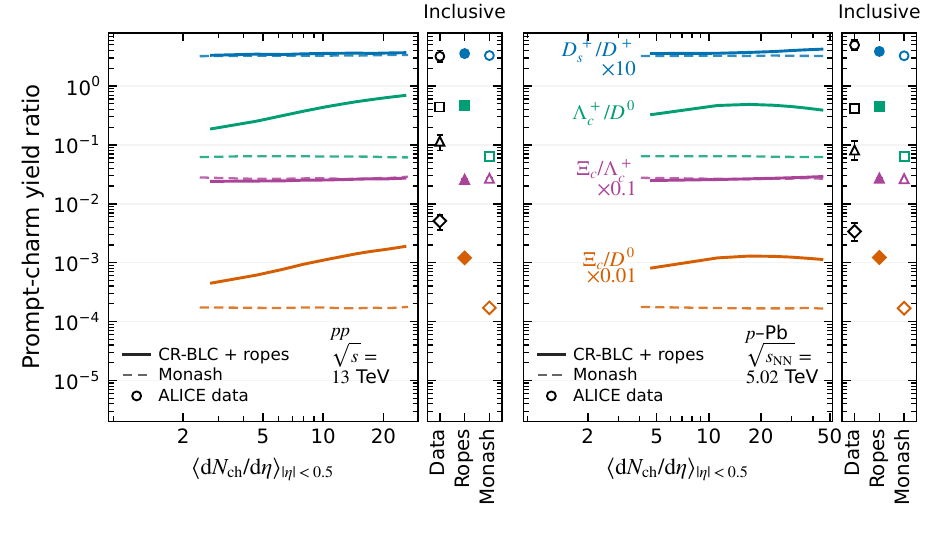}
\caption{Prompt charm ratios in pp at 13 TeV (left) and p--Pb at 5.02 TeV (right). Curves include strong and electromagnetic feed-down, exclude beauty decays, and use $0<p_T<50$ GeV with $|y|<0.5$ (pp) or $-0.96<y_{\rm CM}<0.04$ (p--Pb). Here $\Xi_c=\Xi_c^0+\Xi_c^+$. Side columns compare minimum-bias MC integrals with ratios of ALICE fragmentation fractions~\cite{ALICE:2023sgl,ALICE:2024ocs}, including the experimental low-$p_T$ extrapolation. The summed p--Pb $\Xi_c$ reference assumes equal $\Xi_c^0$ and $\Xi_c^+$ yields.}
\label{fig:ppcharm}
\end{figure}

The inclusive columns of Fig.~\ref{fig:ppcharm} show that the combined prescription brings $\Lambda_c/D^0$ close to the reference values. The inclusive $\Xi_c/D^0$ ratio remains lower by roughly a factor four in pp and three in p--Pb. The apparent $\Xi_c$ deficit remains the main challenge for a full description from light to heavy flavours.

On the experimental side, a reproducible comparison requires an experiment-vetted Rivet implementation of the particle-level analysis, accompanied by the particle definitions, decay and feed-down treatment, beauty subtraction, branching-fraction conversions, low-$p_T$ extrapolations and uncertainty information needed to reconstruct the published results. Measurements of the activity dependence of $\Xi_c/\Lambda_c$, normalized to a reference activity class, could be compared with $\Xi/\Lambda$ to test whether the additional strange quark produces a similar relative rise in the charm and light-baryon sectors.

On the phenomenology side, a dedicated tuning study could test how much of the hadronic $\Xi_c$ discrepancy originates in the deficit already present at the $Z$ pole (Fig.~\ref{fig:zbaseline}). Such a diagnostic tune would prioritize the electron--positron $\Xi_c$ yield, accepting possible deterioration in other species. Its parameters would then be frozen for the extrapolation to hadronic collisions. If this tune described both the baseline yield and the observed multiplicity dependence, it would demonstrate that a baseline correction can remove the discrepancy within the chosen model and support its multiplet extrapolation.

The smaller change in $D_s/D^+$ constrains strangeness in the meson sector. The activity dependence of the charm ratios is a prediction.

\subsection{O--O predictions}\label{sec:nuclear}
\begin{figure}[htbp]
\centering
\includegraphics[width=0.49\linewidth]{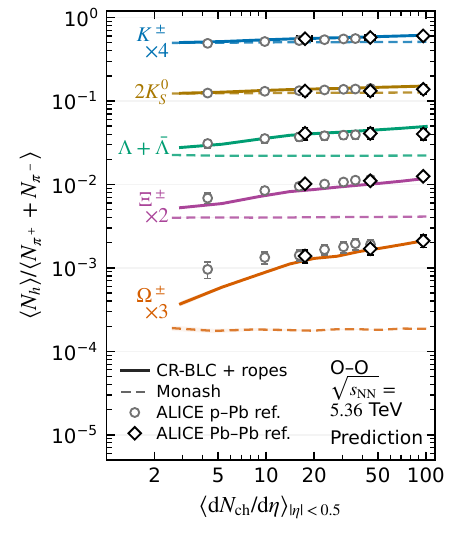}\hfill
\includegraphics[width=0.49\linewidth]{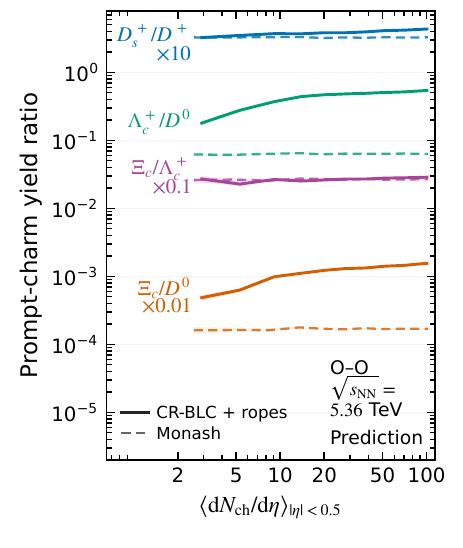}
\caption{O--O predictions at $\sqrt{s_{\rm NN}}=5.36$ TeV with V0M activity classes and $|y|<0.5$. Left: integrated strange-hadron/pion ratios for primary light hadrons; curves and bands follow Fig.~\ref{fig:pp}. ALICE p--Pb and Pb--Pb measurements at 5.02 TeV~\cite{ALICE:2013wgn,ALICE:2015mpp,ALICE:2019hno,ALICE:2025cqy} are used as cross-system references. Right: prompt charm ratios integrated over $0<p_T<50$ GeV; species, decay and line conventions follow Fig.~\ref{fig:ppcharm}.}
\label{fig:nuclear}
\label{fig:oocharm}
\end{figure}

Figure~\ref{fig:nuclear} extends the same prescription to O--O at 5.36 TeV, with p--Pb and Pb--Pb data at 5.02 TeV as cross-system references in the absence of O--O measurements. Comparing the systems at matched multiplicity tests the empirical working hypothesis of an approximately common activity dependence of strangeness enhancement. Their different geometries can produce departures from this trend. The multistrange-baryon/pion ratios respond most strongly to activity, while the kaon/pion ratio changes little. The connected-path criterion lets an extended environment retain additional moving pieces, and saturation moderates their largest assigned breakup forces. The O--O predictions agree satisfactorily with the p--Pb and Pb--Pb reference measurements at comparable multiplicities. The right panel gives O--O predictions for $\Lambda_c/D^0$, $\Xi_c/D^0$, $\Xi_c/\Lambda_c$ and $D_s/D^+$, following the same species and decay definitions as the pp and p--Pb comparison.

\FloatBarrier
\section{Conclusions and next steps}
Junction formation and rope hadronization provide a common microscopic model for changes in baryon formation and strange-quark production, drawing on different channels of the same SU(3) decomposition. The few-parton examples show how these channels arise and which correlations are lost in a leading-colour assignment, while their combination with overlap geometry and fragmentation generates the observed hierarchy of increasing strange-hadron/pion ratios with activity. With common settings, the extension presented here gives a satisfactory description of the light-flavour ratios in pp and p--Pb and supplies O--O predictions. The remaining difficulty with charm, particularly the $\Xi_c$ yield, motivates a dedicated electron--positron baseline tune to determine how much of the discrepancy is inherited from ordinary fragmentation before testing the extrapolation with multiplicity.

The local escape criterion and force saturation introduced here are phenomenological prescriptions, for which the three-string example supplies a possible screening mechanism through background colour correlations. A dynamical calculation should establish the strength and evolution of this screening and determine how it moderates the breakup force as the density increases, building on the more complete treatment of pairwise geometry and relative motion in the parallel-frame construction~\cite{Bierlich:2020naj,Bierlich:2022oja}. Gluon plaquettes and evolving string orientations add subtleties that require a thorough examination of the equations of motion of strings in a dense environment, so that their overlap, colour evolution and fragmentation can be treated consistently.

Hadronic rescattering is the most urgent addition for an extension to larger nuclei, since established transport approaches show that the hadronic stage can change both particle spectra and flavour composition~\cite{Werner:2023jps,Knospe:2015nva,Bierlich:2021poz}. Coupling this stage to interacting strings requires matching the vacuum-based fragmentation construction to the motion and breakup of fields in the dense environment, with consistent hadron production vertices determining where and when subsequent scattering can occur~\cite{Ferreres-Sole:2018vgo,Chakraborty:2022nui}. Developing this matching would allow a quantitative progression towards Pb--Pb.

O--O offers access to the intermediate region between dilute pp events and extended Pb--Pb collisions, where comparisons across systems at similar multiplicity can test the effects of geometry and the lifetime of the interacting system. Higher-luminosity samples will allow more detailed investigations of charm hadronization through the activity dependence of $\Lambda_c/D^0$, $\Xi_c/\Lambda_c$ and $D_s/D^+$, complemented by strangeness balance functions that constrain the accompanying conservation and correlation structure~\cite{Bierlich:2025pkg}. Electron--positron measurements remain essential both for the fragmentation baseline and for the controlled comparison of one and two hadronic $W$ decays, with archived data and future high-statistics $WW$ samples at FCC-ee offering opportunities for matched hyperon measurements. The widening range of collision systems and available statistics will allow increasingly detailed tests of a common microscopic description of colour and flavour.

\section*{Acknowledgements}
Financial support from Vetenskapsrådet contract 2023-04316 is acknowledged.

\bibliographystyle{review}
\bibliography{references}
\end{document}